\documentclass[a4paper,11pt]{article}
\usepackage{epsfig}
\usepackage[utf8]{inputenc}
\usepackage{cite}
\usepackage{hyperref}
\usepackage{epsfig}
\usepackage{amsmath}
\usepackage{amsfonts}
\usepackage{graphicx}
\usepackage{cite}
\usepackage{color}
\usepackage[dvipsnames]{xcolor}
\usepackage{multirow}
\usepackage{amssymb}
\usepackage{graphicx}
\usepackage{caption}

\def\be{\begin{equation}}
\def\ee{\end{equation}}

\newcommand{\bea}{\begin{eqnarray}}
\newcommand{\ena}{\end{eqnarray}}

\begin{document}
\title{{\bf  \LARGE Tidal Love numbers of Proca stars}}
 \author{
{\large Carlos A.~R.~Herdeiro}$^{1}$,\, 
{\large Grigoris Panotopoulos}$^{2}$,\, 
{\large Eugen Radu}$^{1}$
\\
\\
$^{1}${\small Departamento de Matem\'atica da Universidade de Aveiro and } \\ {\small  Centre for Research and Development  in Mathematics and Applications (CIDMA),} \\ {\small    Campus de Santiago, 3810-183 Aveiro, Portugal}
 \\
 \\
$^{2}${\small Centro de Astrof\'isica e Gravitaç\~ao - CENTRA, Departamento de F\'isica,}\\ { \small Instituto Superior T\'ecnico - IST, Universidade de Lisboa - UL,}\\ {\small Av. Rovisco Pais 1, 1049-001, Lisboa, Portugal}
}

\date{June 2020}

\maketitle

\begin{abstract}
Proca stars~\cite{carlos} are everywhere regular, asymptotically flat self-gravitating solitons composed of complex, massive vector bosons, forming a macroscopic, star-sized, Bose-Einstein condensate. They have been suggested as a classical, dynamical model of exotic compact objects, black hole mimickers and dark matter candidates. In spherical symmetry, they are qualitatively similar to their scalar cousins, the standard (scalar) boson stars in all respects studied so far. Here we study the tidal deformability and the quadrupolar tidal Love numbers, both electric-type and magnetic-type, of spherically symmetric  Proca stars. The equations for the perturbations are derived  and the numerical values of the Love numbers for some concrete background solutions are computed. We observe that both Love numbers are  qualitatively similar to the ones of the scalar boson stars; in particular, the electric-type (magnetic-type) quadrupolar Love numbers are positive (negative)  for Proca stars. Quantitatively, for the same compactness, the electric and magnetic Love numbers in the Proca  case are closer (in magnitude) than in the scalar case, with the electric Love numbers having a slightly larger magnitude than the magnetic ones.
\end{abstract}

%\tableofcontents

\newpage

\section{Introduction}

Extracting information on the internal structure of compact  objects, such as neutron stars, is one of the primary goals of current and future gravitational wave (GW) detectors, such as LIGO-Virgo and LISA. The inspiral and relativistic collision of two compact objects in a binary system, and the GW signal emitted during the process, contain a wealth of information on the nature of the colliding bodies. It is expected that ground-based GW detectors may be able to constrain the nuclear equation-of-state (EoS) using the early, low frequency portion of the observed signals from neutron star-neutron star inspirals, such as \textsc{gw170817}~\cite{TheLIGOScientific:2017qsa} and possibly \textsc{gw190425}~\cite{Abbott:2020uma}, or the highly anticipated, but yet to be confirmed, black hole-neutron star inspirals.

\smallskip

In a binary system one of the stars is subjected to the external gravitational field produced by the companion object. The imprint of the EoS within the signals emitted during binary coalescences is mostly determined by adiabatic tidal interactions, characterized in terms of a set of coefficients, known as the \textit{tidal deformability} and the corresponding \textit{tidal Love numbers}. 

\smallskip

The theory of tidal deformability was first introduced in Newtonian gravity over one century ago by Love~\cite{Love1,Love2}, with the purpose of understanding the yielding of the Earth to disturbing forces. For a spherical body, Love introduced two dimensionless numbers to describe the tidal response of the Earth. The first, $h$, describes the relative deformation of the body in the longitudinal direction (with respect to the perturbation); the second, $k$, describes the relative deformation of the gravitational potential. Shortly afterwords,  another dimensionless parameter, $\ell$ was introduced, describing the relative deformation of the Earth in the transverse direction. Observe that the Love number $k$ is not accounting for the extra potential due to the external force; it describes the feedback of the Earth's deformation (due to the external potential) on the overall gravitational potential. 

\smallskip

The consideration of self-gravitating compact objects, such as neutron stars, requires a \textit{relativistic} theory of tidal deformability. This theory was developed in~\cite{flanagan,hinderer,damour,poisson} for spherically symmetric neutron stars and black holes. Naturally, the key deformability parameter is the relativistic generalisation of $k$, since the role of the gravitational potential is played by the metric. Then, tensor perturbations of a spherically symmetric space-time fall into two classes: even parity (or electric, or polar) and odd parity (or magnetic, or axial). Thus, there are electric and magnetic relativistic tidal Love numbers $k^{E}$ and $k^B$, respectively, each of which may, moreover, be decomposed in harmonics of index $l$, with $l$ being the angular degree, thus introducing $k^{E}_l$ and $k^B_l$. Subsequent generalisations to more exotic stars, such as quark and relativistic stars with a polytropic EoS were considered, for instance, in~\cite{lattimer}. The goal of this paper is to consider the relativistic tidal deformability of yet another class of exotic compact objects, namely Proca stars~\cite{carlos}.

\smallskip

According to the latest cosmological data \cite{planck}, non-relativistic matter in the Universe is dominated by dark matter. Despite the success of the concordance cosmological model at large scales ($\gg 1~{\rm  Mpc}$), based on the cold dark matter paradigm, an array of shortcomings at galactic and sub-galactic scales ($<$ a few kpc) persists, such as the core/cusp problem and the missing satellite problem - for reviews see e.g. \cite{review1,review2}. These problems may be tackled if dark matter consists of ultralight scalar particles with a mass $m \ll$ eV~\cite{Hui:2016ltb}.  In these models, ultralight bosons can cluster forming macroscopic Bose-Einstein condensates with the mass of the Sun or even larger. These self-gravitating clumps are called scalar boson stars~\cite{BS1,BS2,BS3,BS4,Harko,Mielke2,BS5,BS6,BS7,BS8}, for spinless bosons, and vector boson stars, or Proca stars~\cite{carlos}, for vector bosons. In particular, some recent proposals  advocate vector bosons can be a dark matter ingredient \cite{vector1,vector2,vector3}. Thus, it becomes interesting, and even phenomenologically relevant, to investigate the potential signatures of these scalar or vector bosonic stars, in particular in relation to GW observations, for which the number of detected events is fast increasing~\cite{ligo1,ligo2,ligo3,ligo4,ligo5}.

\smallskip

So far, the more recently discovered Proca stars qualitatively mimic the scalar boson stars in essentially all studies made for spherically symmetric solutions. Both scalar and vector bosonic stars are perturbatively stable up to some maximal mass~\cite{Gleiser:1988ih,Lee:1988av,carlos}; the unstable solutions share the same three possible fates: migration, collapse or dispersion~\cite{Seidel:1990jh,Guzman:2004jw,Hawley:2000dt,Sanchis-Gual:2017bhw}. As in the scalar case, ultra-compact vector boson stars are unstable against collapse~\cite{Cunha:2017wao}. Both scalar and vector stars can form dynamically, via gravitational collapse of a cloud of scalar/vector bosons due to the mechanism of gravitational cooling~\cite{Seidel:1993zk,DiGiovanni:2018bvo}.  Thus, it is interesting to ask how much this resemblence remains in dynamical, strong gravity systems like binaries, and in their GW phenomenology. One approach to tackle this question is to study such binaries using numerical relativity techniques. Studies in this direction have been reported~\cite{Palenzuela:2006wp,Palenzuela:2007dm,Bezares:2017mzk,Palenzuela:2017kcg,Sanchis-Gual:2018oui,Bezares:2018qwa}. Another approach is to look at the tidal deformability. This is reported in this paper.

\smallskip

For black holes in Einstein's theory, the tidal Love numbers $k^E,k^B$ are precisely zero \cite{poisson}. This intriguing observation means that for a Schwarzschild black hole in an external gravitational field, albeit distorted, its distortion does not feedback on the total gravitational potential. One may say that this potential is infinitely rigid against such deformations. This is not true, however, for exotic compact objects such as scalar boson  stars~\cite{pani}. One may then envisage the following scenario. If this sort of dark matter stars exist, the coalescence and merger of one such binary will have no (considerable) electromagnetic counterpart, unlike a binary where one of the members is a neutron star, mimicking in this respect a  binary black hole merger. Moreover, the endpoint could be a Kerr black hole, mimicking also in this respect the black hole ringdown. But the inspiral would be different, in particular due to the different tidal deformability of boson stars and Schwarzschild black holes. Thus, it is of interest to explicitly analyse the Love numbers of scalar and vector bosonic stars. 

\smallskip

The scalar boson stars Love numbers were studied in~\cite{pani}. It was observed that the electric Love numbers $k^E$ are smaller than the typical ones of neutron stars. In this respect, therefore, the scalar boson stars have a gravitational potential which is more rigid that that of neutron stars, but less rigid that that of black holes. 
The magnetic tidal numbers, $k^B$, on the other hand, are negative, unlike those of neutron stars which are positive. This may be interpreted as a negative feedback effect. That is, the odd parity perturbations produce a deformation of the gravitational potential that works against these deformations. We would like to see if this novel feature remains for vector boson stars. Moreover, it was recently shown that \textit{spinning} vector boson stars are stable and can form dynamically, whereas spinning scalar boson stars are unstable~\cite{Sanchis-Gual:2019ljs}. Thus, in some dynamical aspects, the vector case may be more interesting. As we shall see, in the vector case the electric and magnetic Love numbers are qualitatively similar to those of the scalar case, albeit with some quantitative differences. 

\smallskip

The plan of the manuscript is as follows: After this introduction, in the next section we present the model and the Proca star solutions are reviewed. These serve as the background for the perturbations which are then discussed, the equations for the perturbation being obtained in section three. Our numerical results are shown and discussed in the fourth section, and finally we conclude our work in section five with a discussion and possible implications of our results. We adopt the mostly positive metric signature (-,+,+,+), and we work in natural geometrical units setting $\hbar=c=G=1$.

%%%%%%%%%%%%%%%%%%%%%%%%%%%%%%%
\section{Spherical Proca stars}
%%%%%%%%%%%%%%%%%%%%%%%%%%%%%%%

Let us consider the 4-dimensional Einstein-(complex)Proca theory described by the action
\begin{equation}
S[g_{\mu \nu}, A_\mu] = \int \mathrm{d} ^4x \sqrt{-g} \left[ \frac{R}{16\pi}  - \frac{1}{4} \: F_{\alpha\beta} \bar{F}^{\alpha\beta} - \frac{\mu^2}{2} A_\alpha \bar{A}^\alpha \right],
\end{equation}
where $R$ is the Ricci scalar, $g$ is the determinant of the metric tensor $g_{\alpha\beta}$, $A_\mu$ is the (complex) 1-form potential, $F_{\alpha\beta}=\partial_\alpha A_\beta-\partial_\beta A_\alpha$ is the field strength, $\mu$ is the mass of the Proca field, and an overbar denotes complex conjugation. 

Varying the action with respect to the potential $A_\alpha$ one obtains the Proca field equations 
\begin{equation}
\nabla_\alpha  F^{\alpha \beta} = \mu^2 A^\beta \ .
\end{equation}
Furthermore, varying the action with respect to the metric tensor one obtains Einstein's field equations
\begin{equation}
G_{\alpha\beta}  = 8 \pi T_{\alpha\beta} \ ,
\end{equation}
where $G_{\alpha\beta}$ is the Einstein tensor and $T_{\alpha\beta}$ is the energy-momentum tensor of the Proca field
\begin{equation}
T_{\alpha \beta} = - F_{\sigma ( \alpha} \bar{F}_{\beta )}^\sigma - \frac{1}{4} g_{\alpha\beta} F_{\sigma \tau} \bar{F}^{\sigma \tau} + \mu^2 \left[A_{(\alpha} \bar{A}_{\beta)} - \frac{1}{2} g_{\alpha\beta} A_\sigma \bar{A}^\sigma\right] \ ,
\end{equation}
where $(\alpha,\beta)$ denotes symmetrization with unit weight with respect to the two indices $\alpha,\beta$. 

Following~\cite{carlos},  we seek spherically symmetric solutions by making the usual metric ansatz,
\begin{equation}
ds^2 = -e^{\nu(r)} dt^2 + e^{\lambda(r)} dr^2 + r^2 (d \theta^2 + \sin^2 \theta d \varphi^2) \ .
\label{metrican}
\end{equation}
Alternative, the metric can be parameterised by the two new functions, $m(r),\sigma(r)$, via 
\begin{equation}
e^{-\lambda}  \equiv  N\equiv 1-\frac{2m(r)}{r}  \ , \qquad 
e^\nu   \equiv  \sigma^2 N \ .
\end{equation}
For the Proca field, we make the ansatz
\begin{equation}
\mathcal{A} = e^{-i \omega t} [f(r) dt + i g(r) dr] \ ,
\end{equation}
where $\omega$ is the oscillation frequency of the Proca potential. Observe, however, that the phase $e^{-i \omega t} $ drops out at the level of the energy-momentum tensor, and so does the explicit $t$ dependence, making the Proca ansatz compatible with the metric ansatz above. 
All unknown quantities $N(r), \sigma(r), f(r), g(r)$ are function of the radial coordinate only, and $\omega$ is a real frequency parameter. Then, the Einstein-Proca equations yield the following equations for the metric functions
\begin{eqnarray}
\frac{\sigma'}{\sigma}  & = & 4 \pi r \mu^2 \left(g^2 + \frac{f^2}{\sigma^2 N^2} \right) \ , \\
m'  & = & 4 \pi r^2 \left[ \frac{(f'-\omega g)^2}{2 \sigma^2} + \frac{\mu^2}{2} \left(g^2 N + \frac{f^2}{N \sigma^2} \right) \right] \ ,
\end{eqnarray}
where the prime denotes differentiation with respect to $r$. Additionally, the field equations for the Proca field read
\begin{eqnarray}
\omega g - f'  & = & \frac{\mu^2}{\omega} \sigma^2 N g \ , \\
\frac{\mu^2 r^2 f}{\sigma N}  & = & \frac{d}{dr} \left[ \frac{r^2 (f'-\omega g)}{\sigma} \right] \ .
\end{eqnarray}

\begin{figure*}[ht]
\centering
\includegraphics[width=0.6\textwidth]{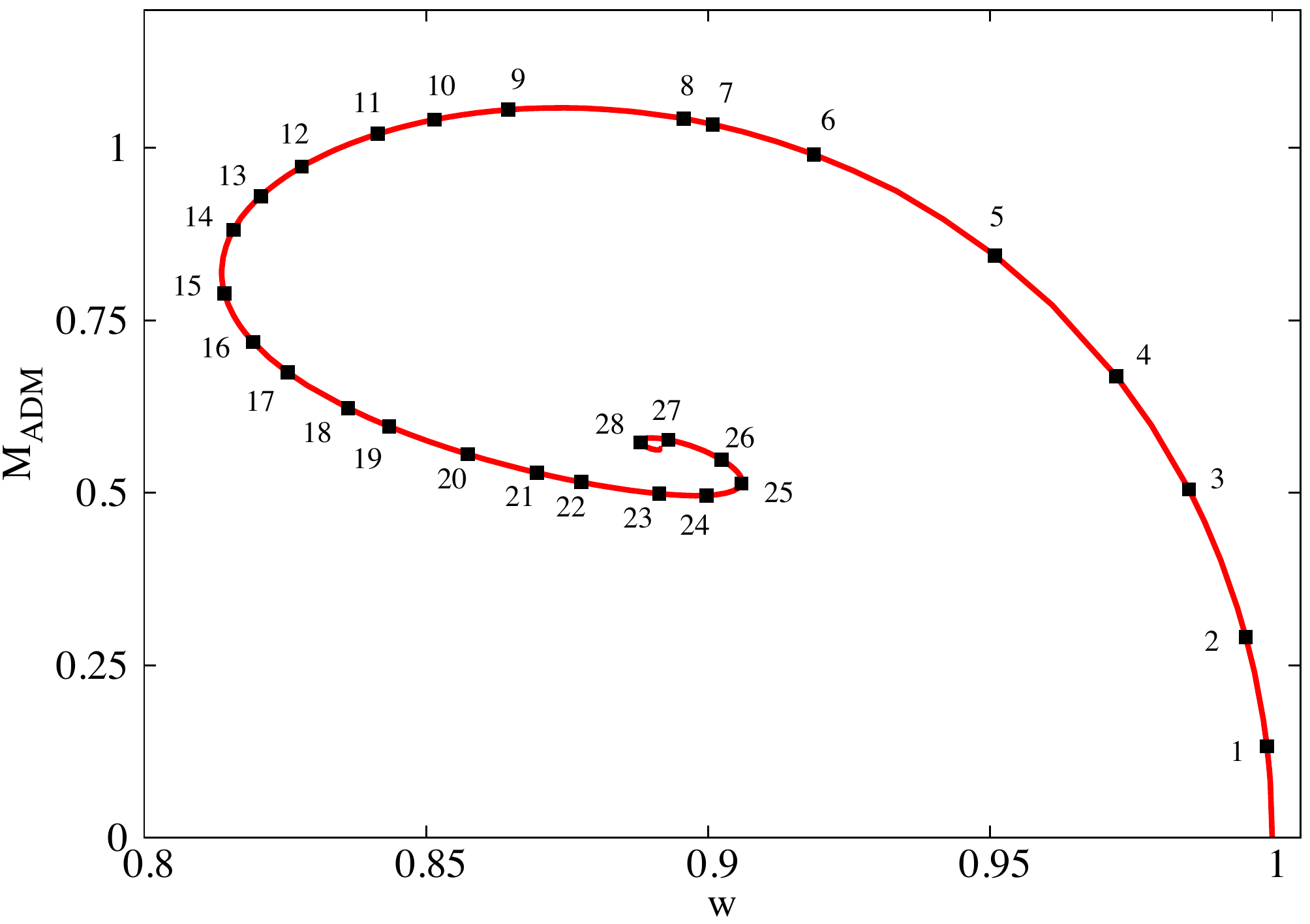}   \
\caption{
Domain of existence of Proca star solutions in an ADM mass $vs.$ Proca field oscillation frequency diagram, both in units of the boson mass $\mu$. Highlighted numbered solutions will be analysed below.
}
\label{fig:1}
\end{figure*}

Proca stars were first obtained in~\cite{carlos} using this framework. Once the solutions have been computed, the ADM mass of the Proca star, $M$ and its ``radius", $R$, defined as the areal radius that encloses 99\% of the mass can be obtained.\footnote{We remark that contrary to neutron stars, Proca stars, like scalar boson stars, do not have a rigid surface. The bosonic field decays exponentially and thus only vanishes at spatial infinity. Nonetheless, the fast fall-off suggests one may define the star radius as one that contains almost all the mass, as above.} Then, we define the \textit{compactness} of the star as
\be
C=\frac{M}{R} \ .
\ee
The domain of existence of the Proca star solutions in a ADM mass $vs.$ Proca frequency diagram is shown in Fig.~\ref{fig:1}. The solutions fall along a spiralling (red solid) curve. The curve starts at $\omega=\mu$ and $M\rightarrow 0$. This is the Newtonian limit in which the stars become very dilute. As one moves up along the spiral, the ADM mass increases and $\omega$ decreases, until a maximal mass is attained. The solutions between this maximal mass solution and the Newtonian limit are perturbatively stable~\cite{carlos}. Beyond the maximal mass, solutions are unstable against perturbations. When perturbed they can have one of three fates: migration to the stable branch, collapse into a black hole or complete fission~\cite{Sanchis-Gual:2017bhw}. In the following, we shall consider 28 solutions marked with black boxes numbered from 1 to 28 in Fig.~\ref{fig:1}. The metric functions $m(r),\sigma(r)$ as well as the vector field components $f(r),g(r)$ for the solution number 5, which has $\omega=0.9508\mu$, are shown in Fig. 2. Notice that $g$ is nodeless, while $f$ changes sign at least once \cite{carlos}. Observe also, from the left panel, that the field profile functions extend to infinity, confirming that Proca stars do not have a rigid surface. This justifies the definition of the stars' radius above.

\begin{figure*}[ht]
\centering
\includegraphics[width=0.47\textwidth]{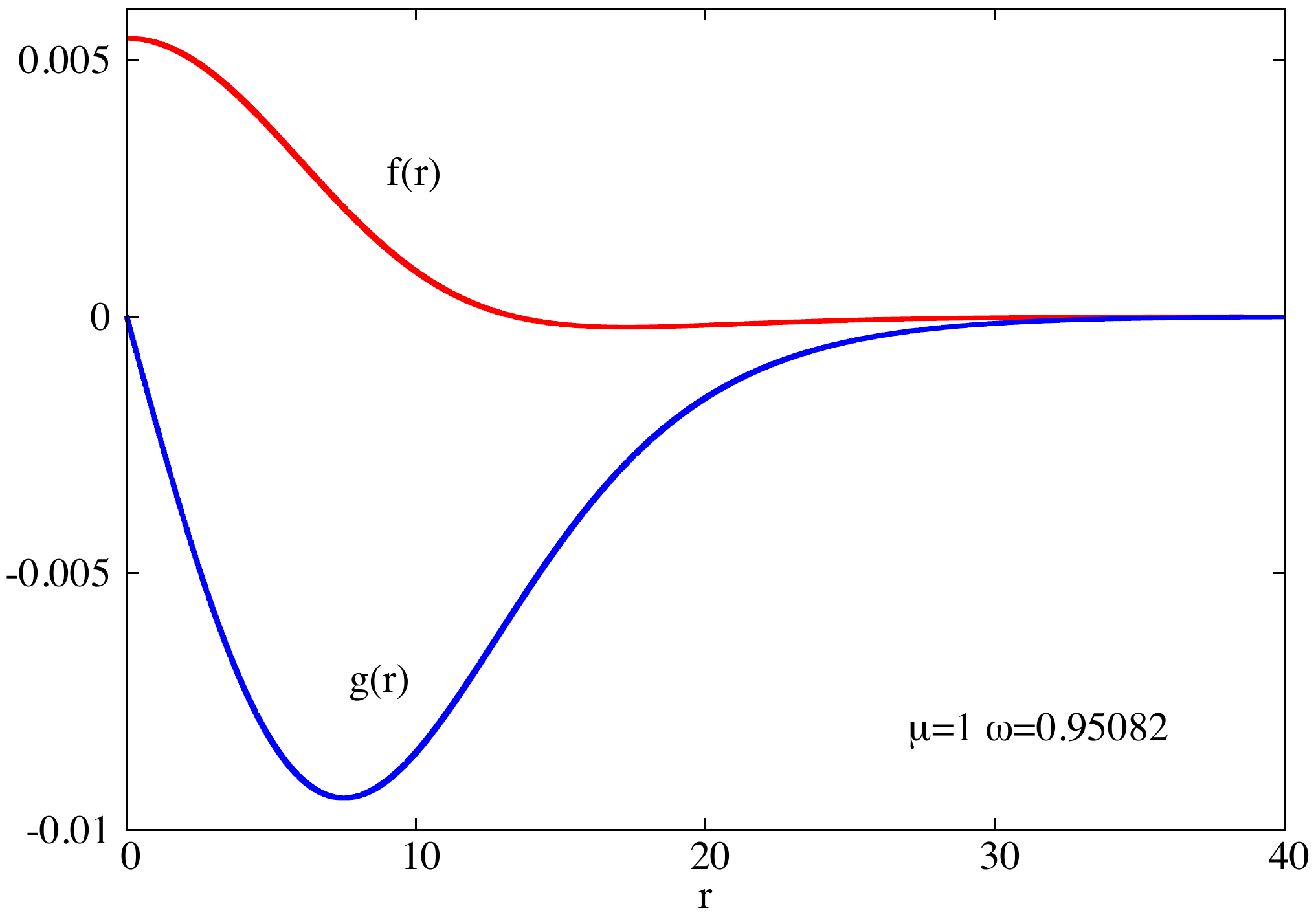}   \
\includegraphics[width=0.47\textwidth]{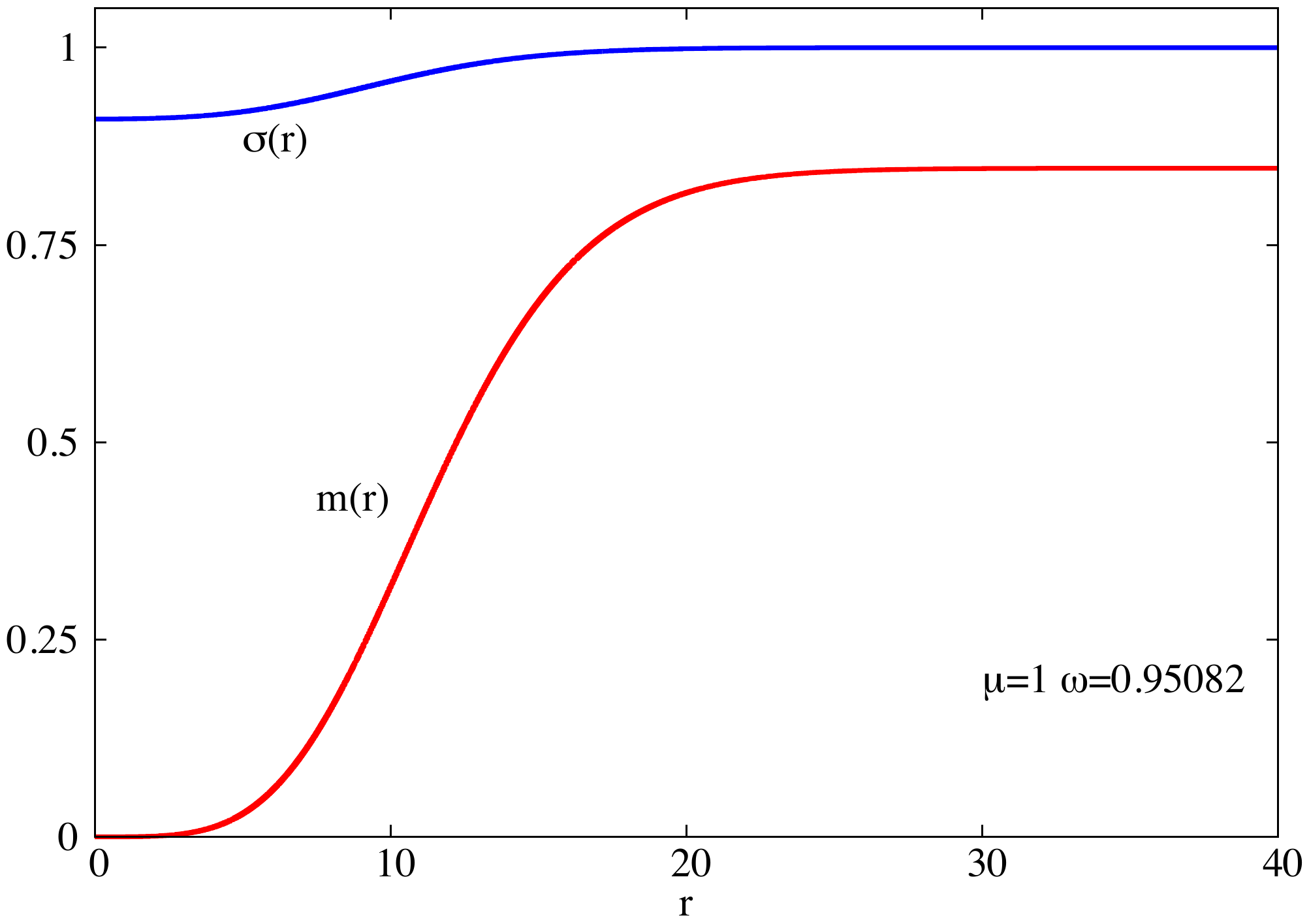}   \
\caption{
		{\bf Left panel:} 
Proca potential components $f(r)$ (red curve) and $g(r)$ (blue curve) versus radial coordinate.
		{\bf Right panel:} 
Metric functions $m(r)$ (red curve) and $\sigma(r)$ (blue curve). Both are for a solution with  $\omega=0.9508\mu$ (solution 5 in Fig.~\ref{fig:1}).
}
\label{fig:2}
\end{figure*}

%%%%%%%%%%%%%%%%%%%%%%%%%%%%%%%%%%%%%%%%%%
\section{Equations for the perturbations}
%%%%%%%%%%%%%%%%%%%%%%%%%%%%%%%%%%%%%%%%%

In order to compute the tidal Love numbers we shall follow the method in~\cite{hinderer,pani}. Thus, the first step is to consider linear perturbations of the Proca star solutions. As such, we perturb the metric as 
\begin{equation}
g_{\mu \nu}  = g_{\mu \nu}^{(0)} + h_{\mu \nu}^{\rm even} + h_{\mu \nu}^{\rm odd} \ ,
\end{equation}
where we explicitly separate even and odd perturbations, and the Proca field as
\begin{equation}
A_{\mu}  = A_{\mu}^{(0)} + \delta A_{\mu} \ ,
\end{equation}
imposing that the perturbations satisfy the linearised Einstein's field equations around the unperturbed background
\begin{equation}
\delta G_{\mu}^{\nu}  = 8 \pi \delta T_{\mu}^{\nu} \ ,
\end{equation}
as well as the linearised  Proca field equations.

The explicit form of the Proca 4-potential perturbations is 
 \cite{Dolan,Berti}
\begin{equation}
\delta A_{\mu} = e^{-i \omega t} \left( \delta A_0(r) Y_l^m, i \delta A_1(r) Y_l^m, i \delta A_2(r) \frac{\partial Y_l^m}{\partial \theta}, \delta A_3(r) \sin \theta \frac{\partial Y_l^m}{\partial \theta} \right) \ ,
\end{equation}
where $Y_l^m$ are the usual spherical harmonics, and the angular degree $l=2$ will be chosen in the explicit computations below, corresponding to quadrupolar deformations. The ``matter" (Proca field) perturbations are therefore characterized by four unknown functions of $r$: $\delta A_0(r),\delta A_1(r),\delta A_2(r),\delta A_3(r)$. 

The metric perturbations, on the other hand, are split into two parts, $h_{\mu \nu}^{\rm even}$ and $h_{\mu \nu}^{\rm odd}$, according to parity. In the Regge-Wheeler gauge the even metric perturbations can be simplified to take the form~\cite{Regge}: 
\begin{equation}
h_{\mu \nu}^{\rm even}dx^\mu dx^\nu  = \left\{ H(r)\left[e^{\nu(r)} dt^2+ e^{\lambda(r)} dr^2\right]+  K(r)r^2\left[d\theta^2+  \sin^2 \theta d\varphi^2\right]\right\} Y_l^m \ ,
\end{equation}
where $\nu,\lambda$ are the unperturbed solution metric functions. Thus, even metric perturbations are characterized by two unknown functions of $r$, $H(r), K(r)$. 

The explicit form of the metric perturbations for the odd case, can be simplified to take the form \cite{pani}, 
\be
h_{\mu \nu}^{\rm odd}dx^\mu dx^\nu  = h(r)\left\{  -\frac{1}{\sin\theta}\frac{\partial Y_l^m}{\partial \varphi}d\theta + \sin\theta \frac{\partial Y_l^m}{\partial \theta} d\varphi \right\} dt \ ,
\ee
and it contains a single unknown function $h(r)$.

\smallskip

Since the odd metric perturbations couple to $\delta A_3$ only, while the even metric perturbations only couple to $\delta A_0(r),\delta A_1(r),\delta A_2(r)$, in the following we tackle the two types of perturbations with different parities as two separate problems. First, we solve the system of coupled even perturbations to compute the electric-type Love numbers $k_2^E$. Then, we solve the system of coupled odd perturbations to compute the magnetic-type Love numbers $k_2^B$.

We remark that for a given angular degree $l$, any perturbation $G(x^\mu)$ may be written in the following form
\begin{equation}
G(t,r,\theta,\varphi) = \sum_{m=-l}^{m=l} e^{-i \omega t} G_l(r) Y_l^m(\theta,\varphi) \ ;
\end{equation}
where in the sum $m$ takes $(2l+1)$ values, $m=-l,...,l$, and the unknown radial part $G_l(r)$ satisfies an ordinary differential equation that does not depend on $m$. Therefore, in the following for simplicity, we set $m=0$, and as already mentioned, we focus on the $l=2$ case for the computation of the quadrupolar tidal Love numbers.

%%%%%%%%%%%%%%%%%%%%%%%%%%%%%%%%
\subsection{Even perturbations}
%%%%%%%%%%%%%%%%%%%%%%%%%%%%%%%%

To obtain the perturbation equations in the even case, we use the $r\theta$ and the $rr$ Einstein equations to express $K(r), K'(r)$ in terms of the other metric perturbation $H(r)$. Then the $tt$ field equation leads to a differential equation for $H(r)$ coupled to the perturbations of the Proca field, which reads
\begin{equation}
H''(r) + c_1(r) H'(r) + c_0(r) H(r) + Q(\delta A_0,\delta A_1, \delta A_2)  = 0 \ ,
\label{Heq}
\end{equation}
where the coefficients $c_0, c_1$ are certain functions of the background quantities; explicitly, they are found to be
\begin{eqnarray}
c_1 & = & \frac{2}{r} + \frac{\nu'-\lambda'}{2} \ , \\
c_0 & = & -6 \frac{e^\lambda}{r^2} + \frac{\lambda' + 3 \nu'}{r} - \frac{\lambda' \: \nu'}{2} -\frac{(\nu')^2}{2} + \nu'' + 16 \pi \mu^2 [e^{\lambda - \nu} \: f^2] \ ,
\end{eqnarray}
with $Q$ being a complicated expression of the vector field perturbations $\delta A_i, i=0,1,2$, which we shall omit. It turns out that the equation for $\delta A_1$ is an algebraic equation, which can then be used to express $\delta A_1$ in terms of $\delta A_0$ and $\delta A_2$:
\begin{equation}
\delta A_1 = \frac{r^2 \omega \delta A_0'-r^2 \mu^2 e^\nu H g-6 e^\nu \delta A_2'}{r^2 (\omega^2-\mu^2 e^\nu)-6 e^\nu}  \ .
\label{evena1}
\end{equation}

The two remaining equations for the Proca field, for the components $\delta A_0,\delta A_2$, have a with similar structure. For instance, the one for $\delta A_2$, which is simpler, is found to be
\begin{equation}
\delta A_2'' + \frac{\nu'-\lambda'}{2} \delta A_2' 
+ e^\lambda (-\mu^2+\omega^2 e^{-\nu}) \delta A_2 
+ \frac{1}{2}{(\delta A_1 (\lambda'-\nu')-2 \delta A_1'-2 \omega \delta A_0 e^{\lambda-\nu})} = 0 \ .
\label{evena2}
\end{equation}

%%%%%%%%%%%%%%%%%%%%%%%%%%%%%%%
\subsection{Odd perturbations}
%%%%%%%%%%%%%%%%%%%%%%%%%%%%%%

For  the odd case, the Einstein equations and the $\delta A_3$ Proca equation yield a system of two coupled equations for $h(r),\delta A_3(r)$, which is the following
\begin{equation}
h'' - \frac{\lambda' + \nu'}{2} \: h' + \frac{r (\lambda' + \nu')-4 e^\lambda-2}{r^2} \: h + 16 \pi \left[\mu^2 e^\lambda f \delta A_3 -  (\omega g - f') \delta A_3'\right] = 0 \ ,
\label{hodd}
\end{equation}
for $h(r)$, and
\begin{equation}
\delta A_3'' + \frac{\nu'-\lambda'}{2} \: \delta A_3' + \frac{e^\lambda (-r^2 \mu^2+r^2 \omega^2 e^{-\nu}-6)}{r^2} \: \delta A_3 + \frac{e^{-\nu}}{2} [2 h' (f'-\omega g) + h \: \chi] = 0 \ ,
\label{odda3}
\end{equation}
for $\delta A_3$, where $\chi$ is found to be
\begin{equation}
\chi = 2 (-\mu^2 e^\lambda f + f'' - \omega g') + (\lambda' + \nu') (w g - f') \ .
\end{equation}

%%%%%%%%%%%%%%%%%%%%%%%%%%%%%%%%%%%%%%%%%%%%%%%%%%%%%%%%%%%%
\section{Computation of the quadrupolar tidal Love numbers}
%%%%%%%%%%%%%%%%%%%%%%%%%%%%%%%%%%%%%%%%%%%%%%%%%%%%%%%%%%%%

Having computed the equations for the linearised  perturbations of the Einstein-Proca model, we can now establish the methodology to compute the tidal Love numbers. This follows other cases in the literature. Nonetheless, to make our paper self-contained, we briefly explain here the method.

\smallskip

First, regarding even perturbations and electric-type Love numbers we follow closely \cite{hinderer}. Although Proca stars do not have a surface, as discussed above, we shall assume that outside what we defined as the radius of the star, we are approximately in vacuum. Then, outside the star, $\delta A_\mu = 0 = A_\mu$. By Birkshoff's theorem, the background metric components outside the star correspond to those of the Schwarzschild solution, $e^{\nu(r)}=1-2M/r=e^{-\lambda(r)}$. Consequently, the equation for $H$, eq.~\eqref{Heq}, takes the simple form
\begin{equation}
H''(r) + 2 \frac{r-M}{r (r-2 M)} H'(r) -2 \frac{2M^2-6 M r+3 r^2}{r^2 (r-2 M)^2} H(r) = 0 \ .
\label{Heq2}
\end{equation}
Setting $x\equiv -1+r/M$, eq. \eqref{Heq2} can be recast as the associated Legendre equation with $l=2=m$. The solution of the latter is given by
\begin{equation}
H(x) = C_2 L_2^2(x) + C_1 Q_2^2(x) \ ,
\label{Hr}
\end{equation}
where $C_1,C_2$ are two arbitrary coefficients, $L_2^2(x)$ is the standard Legendre polynomial and $Q_2^2(x)$ is the Legendre function of the second kind. 

\smallskip

Now we consider the Proca star with mass $M$ and radius $R$ to be subjected to an external gravitational field, produced, for instance, by a companion star in a binary, see $e.g.$ Fig. 1 in~\cite{GuerraChaves:2019foa}. The Proca star will react to the external field by deforming. The leading deformation is to develop a quadrupolar moment $Q_{ij}$, which is proportional to the static external quadrupolar tidal field $\mathcal{E}_{ij}$ 
\begin{equation}
Q_{ij} = - \lambda \mathcal{E}_{ij} \ .
\end{equation}
The constant $\lambda$ is called the ($l=2$) \textit{tidal deformability}. It relates to the dimensionless tidal Love number $k$, by definition, as
\begin{equation}
\lambda = \frac{2}{3} k_2 R^5 \ .
\end{equation}
One can also introduce the \textit{dimensionless deformability} as follows
\begin{equation}
\Lambda = \frac{2}{3} \frac{k_2}{C^5}=\frac{\lambda}{M^5} \ .
\end{equation}

\smallskip

To compute $k_2$ one proceeds as follows. At large distances, the Newtonian gravitational potential of the full system (Proca star plus perturbation) reads
\begin{equation}
\frac{-1-g_{tt}(r \rightarrow \infty)}{2} = \Phi(r \rightarrow \infty) \sim - \frac{M}{r} - \frac{3 Q_{ij}}{2 r^3} \left(n^i n^j-\frac{1}{3} \delta^{ij}\right)+\frac{1}{2} \mathcal{E}_{ij} x^i x^j + ... \ , 
\label{Npot}
\end{equation}
where $n^i\equiv x^i/r$ are the components of the unit 3-vector, and the dots indicate higher order terms that are neglected.

\smallskip

Using the asymptotic behaviour for $H(r)$ obtained from eq.~\eqref{Hr} and comparing with~\eqref{Npot}, the coefficients $C_1,C_2$ are found to be
\begin{eqnarray}
C_1 & = & \frac{15 \lambda \mathcal{E}}{8 M^3} \ , \\
C_2  & = &  \frac{M^2 \mathcal{E}}{3} \ .
\end{eqnarray}
Here $\mathcal{E}$ is the $l=2=m$ coefficient of an expansion of $ \mathcal{E}_{ij}$ in a basis of symmetric traceless tensors - see~\cite{hinderer,Thorne:1980ru}.

\smallskip

Knowing $C_1,C_2$, as function of $k_2^E$ (say), replacing back on eq.~\eqref{Hr}, one computes a new function 
\be
y(r,k_2^E) \equiv r \frac{H'(r)}{H(r)} \ .
\ee
Then one fixes a certain radius at which the exterior solution is valid. For a neutron star this could be the hard surface. For the case of a Proca star we choose a large radius $r_{\rm L}$. Let $y\equiv y(r_{\rm L},k^E_2)$.  Solving this equation for $k^E_2$ one obtains the Love number as a function of $y$ and the compactness $C$:
\begin{equation}
k_2^E = \frac{8C^5}{5} \: \frac{(1-2C)^2 \: [2 C (y-1)-y+2]}{3 \: (1-2 C)^2 \: [2 C (y-1)-y+2] \: ln(1-2C) + P_5(C)} \ ,
\label{elove}
\end{equation}
where $P_5(C)$ is a fifth order polynomial given by
\begin{equation}
P_5(C) = 2 C \: [4 C^4 (y+1) + 2 C^3 (3 y-2) + 2 C^2 (13-11 y) + 3 C (5 y-8) -3 y + 6] \ .
\end{equation}
Formula~\eqref{elove} determines the quadrupolar electric Love number knowing $y$ and the compactness of the star.

\smallskip

Let us now turn our attention to the odd perturbations and magnetic Love numbers. Outside the star the equation for $h$, eq.~\eqref{hodd}, becomes
\begin{equation}
h'' - \frac{\lambda' + \nu'}{2} \: h' + \frac{r (\lambda' + \nu')-4 e^\lambda-2}{r^2} \: h  = 0 \ ,
\end{equation}
where the metric functions $\nu,\lambda$ are the ones of  the usual Schwarzschild solution. Using the explicit form for these functions, $h(r)$ satisfies the simple equation
\begin{equation}
h'' + \frac{4M-6r}{r^2 (r-2M)} \: h = 0 \ ,
\end{equation}
which can be solved in terms of elementary functions. Going through the same steps as for the even case, one obtains the formula valid for neutron stars and boson stars found in e.g. \cite{damour,pani}. One obtains the following expression for the magnetic Love numbers
\begin{equation}
k_2^B = \frac{8C^5}{5} \: \frac{2 C (y-2)-y+3}{3 \: [2 C (y-2)-y+3] \: ln(1-2C) + P_4(C)} \ ,
\label{blove}
\end{equation}
where $P_4(c)$ is a fourth order polynomial given by
\begin{equation}
P_4(C) = 2 C \: [2 C^3 (y+1) + 2 C^2 y +3 C (y-1) -3 y + 9] \ .
\end{equation}
Formula~\eqref{blove} determines the quadrupolar magnetic Love number knowing $y$ and the compactness of the star.

\smallskip

We remark that both~\eqref{elove} and~\eqref{blove} are the same as in~\cite{hinderer,pani}. Although the matter fields are different from those in~\cite{hinderer,pani}, and consequently the structure of the perturbation equations is also different (and more involved in our case),  sufficiently far away the equations for the perturbations take the same form, hence leading to the same end results~\eqref{elove} and~\eqref{blove}.

%%%%%%%%%%%%%%%%%%
\section{Results}
%%%%%%%%%%%%%%%%%%

At this point, it remains to compute $y$ for any given solution, since $C$ is known for that solution. To compute $y$ we integrate the full system of coupled perturbations. For the even case, these are equations~\eqref{Heq},~\eqref{evena2} and the corresponding equation for $\delta A_0$, using also~\eqref{evena1}. For the odd case these are equations~\eqref{hodd} and~\eqref{odda3}. Analysing the solution of the perturbation equations near the origin, one obtains $y(r=0)=2$ for even perturbations and  $y(r=0)=3$ for odd perturbations. These values are used to initialise the radial integration in order to obtain the value of $y=y(r_{\rm L})$.  This value is independent of the initial value taken for the matter field perturbations (i.e. the value at the origin), as long as $y$ is sufficiently asymptotic and the initial value of the matter perturbations is compatible with the asymptotic decay of these perturbations. 

\begin{figure*}[ht]
\centering
\includegraphics[width=0.485\textwidth]{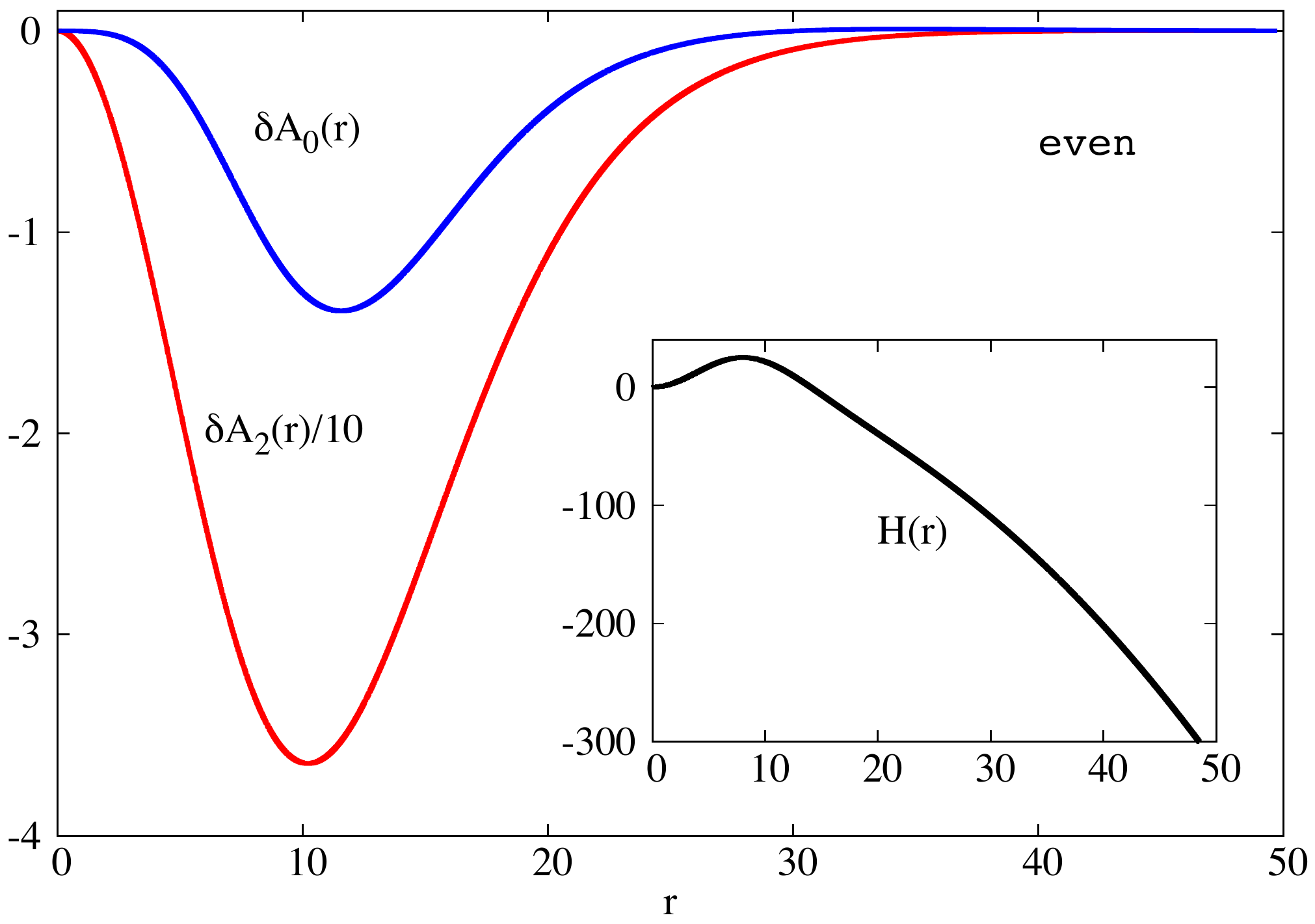}   \ \
\includegraphics[width=0.485\textwidth]{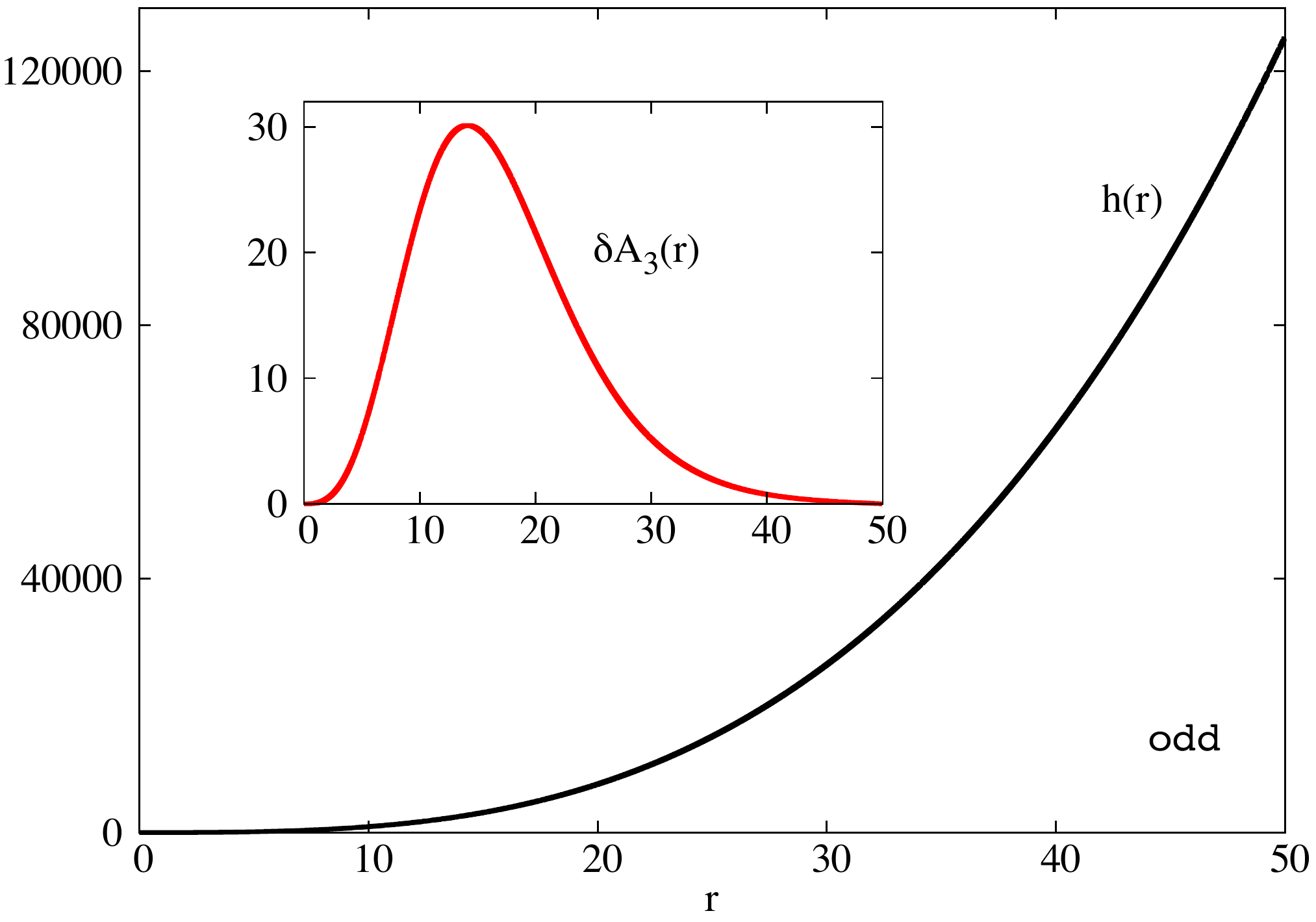}   \
\caption{
Typical form of the perturbation functions for the electric	({\bf left panel}) and magnetic  ({\bf right panel}) cases. The background Proca star is solution 5.
}
\label{fig:32}
\end{figure*}

In practice, the perturbation equations are solved by using a standard Runge–Kutta ordinary differential equations solver and implementing a shooting method in terms of the parameters
$a_0,a_2,a_3$ which determine the approximate expansion of the matter perturbations at $r=0$, 
with 
\be
\delta A_0 = a_0 r^2+\mathcal{O}(r^4) \ , \qquad 
\delta A_2 = a_2 r^2+\mathcal{O}(r^4) \ , \qquad 
 \delta A_3 = a_3 r^2+\mathcal{O}(r^5) \ ,
\ee
such that the $\delta A_0$,  $\delta A_2$ and  $\delta A_3$ vanish asymptotically.
The profile of typical metric and matter perturbations 
are shown in Fig.~\ref{fig:32}, wherein  the background 
Proca star corresponds  to the  solution 5 (which is displayed in Fig.~\ref{fig:2}).

%%%%%%%%%%%%%%%%%FIGURE%%%%%%%%%%%%%%%%%%%%%%%%%%%%%%%%

%%%%%%%%%%%%%%%%%FIGURE%%%%%%%%%%%%%%%%%%%%%%%%%%%%%%%%

\begin{figure*}[ht]
\centering
\includegraphics[width=0.485\textwidth]{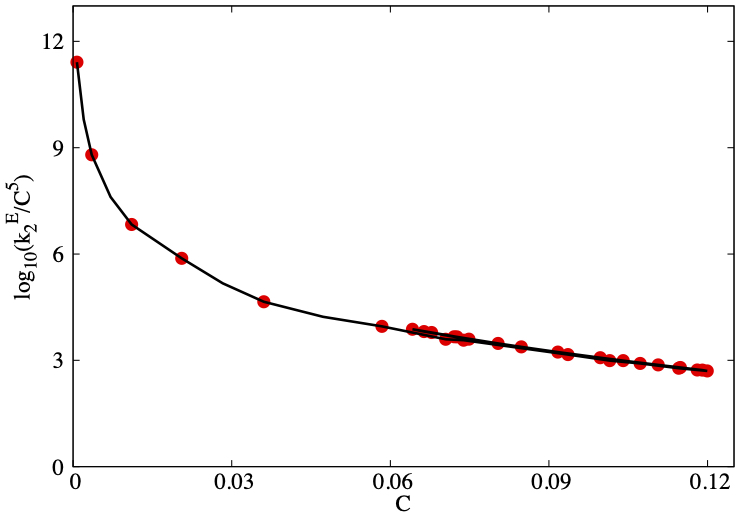}   \ \
\includegraphics[width=0.485\textwidth]{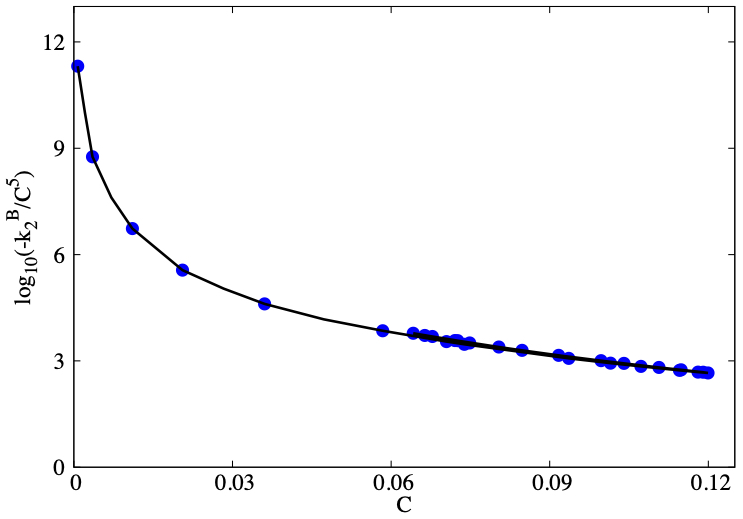}   \
\includegraphics[width=0.5\textwidth]{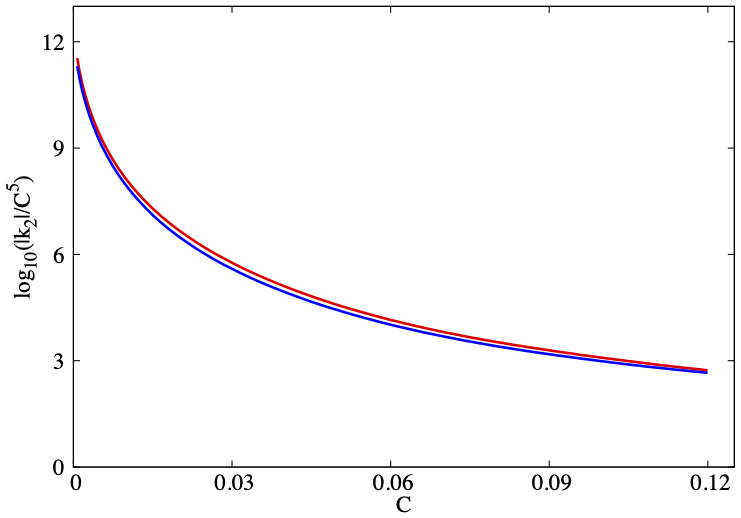}
\caption{Tidal Love numbers of Proca stars $vs.$ compactness (logarithmic plots). The fitting curves are shown as well for the top panels.
		{\bf Top left panel:} 
Electric tidal Love numbers.
		{\bf Top right panel:} 
Magnetic tidal Love numbers. 
{\bf Bottom  panel:} 
Both electric (upper curve in red) and magnetic (lower curve in blue) Love numbers (interpolated curves).
}
\label{fig:3}
\end{figure*}

Our main numerical results are  shown in Fig.~\ref{fig:3}  and also summarized in Table 1 below. In particular, in  the top panels of Fig.~\ref{fig:3} we exhibit the electric (left panel) and the absolute value of magnetic (right panel) Love numbers (divided by $C^5$) of Proca stars versus compactness $C$ in logarithmic plots.  As for the case of scalar boson stars, the electric is positive and the magnetic is negative, or neutron stars, in which case both are positive. Thus, on the $y$ axis we plot $k_2^E$ and/or $|k_2^{B}|$. We fit the numerical results for both cases with a smooth function (a sum of negative power-laws), which read
\be
\log(k_2^E/C^5) = -\frac{3.11}{C^{0.3}} + \frac{14.89}{C^{0.2}} - \frac{11.37}{C^{0.1}} \ ,   \qquad  \log(|k_2^B|/C^5) = -\frac{4.62}{C^{0.3}} + \frac{19.2}{C^{0.2}} - \frac{14.47}{C^{0.1}} \ ,
\ee
which is shown as well. In the bottom panel we exhibit on the same plot both electric-type and (minus) magnetic-type tidal Love numbers $vs.$ compactness (omitting the data from the second branch). One can observe that the electric Love numbers are slightly larger than the magnetic-type ones; thus, the red curve  (corresponding to $k_2^E$) lies slightly above the blue one (corresponding to $|k_2^B|$).

%%%%%%%%%%%%%%%%%FIGURE%%%%%%%%%%%%%%%%%%%%%%%%%%%%%%%%

%%%%%%%%%%%%%%%%%%FIGURE%%%%%%%%%%%%%%%%%%%%%%%%%%%%%%%%%%%%%%%

%%%%%%%%%%%%%%%%%%%%%%%%%%%%%%%%%%%%%
\section{Conclusions and discussion}
%%%%%%%%%%%%%%%%%%%%%%%%%%%%%%%%%%%%%

In this work we have computed the quadrupolar deformability and the corresponding tidal Love numbers, both electric-type $k_2^E$ and magnetic type, $k_2^B$, of spherically symmetric (non-rotating) bosonic stars composed of a complex Proca field. First, the system of coupled perturbations both for the metric and the vector boson were derived. Then, using the known background solution for Proca stars~\cite{carlos}, we integrated the equations of the perturbations under suitable boundary conditions to obtain the numerical values of the quadrupolar Love numbers, both magnetic and electric, as well as the dimensionless (electric) deformability, for a sample of illustrative solutions corresponding to different compactness of Proca stars. Our main numerical results are summarized in Table 1, and for better visualization we have shown them graphically, see Fig.~\ref{fig:3}.

\smallskip

Our main conclusion is that the tidal Love numbers of Proca stars are qualitatively similar to those of the scalar boson stars. Thus, also in this case, the spherical scalar and vector stars parallel each other. Thus, the electric (magnetic) Love numbers of Proca stars are positive (negative), as for scalar boson stars. We recall that for neutron stars all Love numbers are positive. This indicates a positive (negative) feedback effect under an external quadrupolar field, for even (odd) metric perturbations. The full implications of such result in terms of the GW signal in a binary of Proca stars are not, however, fully clear. We have also observed that, in magnitude, the Love numbers are simlar for Proca stars and for scalar boson stars, for the same compactness $C$ - compare, for instance, Fig.~\ref{fig:3} above and Fig.~2 in~\cite{Franzin:2017mtq}. As a noticeable quantitative difference, $k_2^E$ and  $|k_2^B|$ are closer for Proca stars than for scalar boson stars, but in both cases, the electric Love numbers are larger than the magnetic ones for Proca stars. 

\smallskip

Our work suggests there are differences in the GW signal of a binary of spherical bosonic stars with the same compactness, when comparing the scalar and vector cases, \textit{during the inspiral}. This differences may be revealed by more detailed studies of these systems, using numerical relativity techniques. It would be interesting to  perform such studies, especially in view of the ongoing and future science runs of the LIGO-Virgo detectors.

%%%%%%%%%%%%%%%%%%%%%%%%%%%%%%%%%%%%%%%%%%%%%%%%%%%%%%%%%%%%%%%%%%%%%%%%%%

\section*{Acknowlegements}

It is a pleasure to thank V.~Cardoso and E.~Franzin for discussions and correspondence. 
This  work  is  supported  by  the Center  for  Research  and  Development  in  Mathematics  and  Applications  (CIDMA)  through  the Portuguese Foundation for Science and Technology (FCT - Funda\c c\~ao para a Ci\^encia e a Tecnologia), references UIDB/04106/2020 and UIDP/04106/2020, by the Center for Astrophysics and Gravitation - CENTRA, Instituto 
Superior T\'ecnico, Universidade de Lisboa,  through the Grant No. UIDB/00099/2020 and by national funds (OE), through FCT,I.P., in the scope of the framework contract foreseen in the numbers 4, 5 and 6 of the article 23,of the Decree-Law 57/2016, of August 29, changed by Law 57/2017, of July 19.  We acknowledge support  from  the  projects  PTDC/FIS-OUT/28407/2017  and  CERN/FIS-PAR/0027/2019.   This work has further been supported by the European Union’s Horizon 2020 research and innovation (RISE) programme H2020-MSCA-RISE-2017 Grant No. FunFiCO-777740.  The authors would like to acknowledge networking support by the COST Action CA16104.

%%%%%%%%%%%%%%%%%%%%%%%%TABLE%%%%%%%%%%%%%%%%%%%%%%%%%%%%%%%%%%%%%%%%%%%%%%%%%%

%\appendix

\begin{table}
\begin{center}
\begin{tabular}{l | l l l l l l}
  & \multicolumn{6}{c}{{\sc Properties of Proca stars}} \\
 $\#$ solution & $\omega$ & $M$ & $R$ & $C=M/R$ & $k_2^E/C^5$ & $-k_2^B/C^5$  \\
\hline
\hline
1 & 0.99905 & 0.13241 & 184.41730  & 0.00072 & $2.58 \times 10^{11}$ & $2.07 \times 10^{11}$ \\
2 & 0.99533 & 0.29076 & 82.69550  & 0.00352 & $6.29 \times 10^8$  & $5.66 \times 10^8$ \\
3 & 0.98521 & 0.50501 & 45.63673  & 0.01107 & $6.79 \times 10^6$ & $5.44 \times 10^6$  \\
4 & 0.97234 & 0.66911 & 32.57765  & 0.02054 & $7.55 \times 10^5$  & $3.63 \times 10^5$ \\
5  & 0.95082 & 0.84407 & 23.40301 & 0.03607 & 45026.60 & 40590.0  \\
6 & 0.91875 & 0.99050 & 16.95510 & 0.05842 & 9120.43  & 7078.61  \\
7  & 0.90080 & 1.03397 & 14.67271 & 0.07047 & 3935.94  & 3502.5  \\
8 & 0.89564 & 1.04227 & 14.11078 & 0.07386 & 3696.34  & 2931.16  \\
9 & 0.86457 & 1.05538 & 11.27512 & 0.09360 & 1450.48  & 1182.42  \\
10  & 0.85144 & 1.04100 & 10.25753 & 0.10149 & 983.19  & 864.306  \\
11 & 0.84146 & 1.02034 & 9.51483  & 0.10724 & 820.63  & 697.916  \\
12 & 0.82793 & 0.97287 & 8.49397  & 0.11454 & 600.16  & 541.069  \\
13  & 0.82067 & 0.93003 & 7.87919  & 0.11804 & 532.40  & 482.704  \\
14 & 0.81582 & 0.88133 & 7.34898  & 0.11993 & 502.12  & 456.211  \\
15 & 0.81426 & 0.78899 & 6.62957  & 0.11901 &  529.22 & 477.85  \\
16 & 0.81931 & 0.71867 & 6.25762  & 0.11485 & 627.07  & 559.022  \\
17 & 0.82549 & 0.67444 & 6.09488  & 0.11066 & 743.94 & 655.547  \\
18  & 0.83613 & 0.62290 & 5.98784  & 0.10403 & 982.69  & 850.961  \\
19 & 0.84343 & 0.59612 & 5.97916  & 0.09970 & 1186.7 & 1016.55  \\
20 & 0.85737 & 0.55596 & 6.06367  & 0.09169 & 1711.91 & 1438.41   \\
21 & 0.86964 & 0.52909 & 6.24049  & 0.08478 & 2398.54 & 1984.0  \\
22  & 0.87750 & 0.51537 & 6.41172  & 0.08038 & 3008.39  & 2464.83  \\
23 & 0.89134 & 0.49865 & 6.86960  & 0.07259 & 4609.32  & 3717.86  \\
24 & 0.89969 & 0.49611 & 7.31668  & 0.06781 & 6092.81 & 4873.33  \\
25 & 0.90590 & 0.51336 & 7.99704 & 0.06419 & 7551.27  & 6014.75  \\
26  & 0.90236 & 0.54805 & 8.25666  & 0.06638 & 6501.28  & 5212.29  \\
27  & 0.89293 & 0.57683 & 8.00467  & 0.07206 & 4612.79  & 3740.73  \\
28  & 0.88805 & 0.57290 & 7.65207  & 0.07487 & 3953.22  & 3219.83  \\
\end{tabular}
\caption{Frequency, mass, radius, compactness and tidal Love numbers (divided by $C^5$) of Proca stars for the 28 solutions considered here (setting $\mu=1$).} 
\end{center}
\label{table:firstset}
\end{table}

%%%%%%%%%%%%%%%%%%%%%%TABLE%%%%%%%%%%%%%%%%%%%%%%%%%%%%%%%%%%%%%%%%%%%%

\begin{table}
\begin{center}
\begin{tabular}{l | l l l}
  & \multicolumn{3}{c}{{\sc Tidal Love numbers of Proca stars}} \\
 $\#$ solution & $C=M/R$ & $k_2^E$ & $-k_2^B$  \\
\hline
\hline
1   & 0.00072  & 0.00005  & 0.00004 \\
2   & 0.00352  & 0.00034  & 0.00031 \\
3   & 0.01107  & 0.00113  & 0.00090 \\
4   & 0.02054  & 0.00276  & 0.00133 \\
5   & 0.03607  & 0.00275  & 0.00248  \\
6   & 0.05842  & 0.00621  & 0.00482 \\
7   & 0.07047  & 0.00684  & 0.00609  \\
8   & 0.07386  & 0.00812  & 0.00644 \\
9   & 0.09360  & 0.01042  & 0.00849  \\
10  & 0.10149  & 0.01059  & 0.00931  \\
11  & 0.10724  & 0.01164  & 0.00990 \\
12   & 0.11454 & 0.01183  & 0.01067 \\
13   & 0.11804 & 0.01220  & 0.01106 \\
14   & 0.11993 & 0.01246  & 0.01132 \\
15   & 0.11901 & 0.01263  & 0.01141  \\
16   & 0.11485 & 0.01253  & 0.01117 \\
17   & 0.11066 & 0.01235  & 0.01088  \\
18   & 0.10403 & 0.01197  & 0.01037 \\
19   & 0.09970 & 0.01169  & 0.01001 \\
20   & 0.09169 & 0.01109  & 0.00932  \\
21   & 0.08478 & 0.01051  & 0.00869 \\
22   & 0.08038 & 0.01009  & 0.00827 \\
23   & 0.07259 & 0.00929  & 0.00749 \\
24   & 0.06781 & 0.00874  & 0.00699 \\
25   & 0.06419 & 0.00823  & 0.00655  \\
26   & 0.06638 & 0.00838  & 0.00672 \\
27   & 0.07206 & 0.00896  & 0.00727  \\
28   & 0.07487 & 0.00930  & 0.00757 \\
\end{tabular}
\caption{Compactness and quadrupolar tidal Love numbers of Proca stars for the 28 solutions considered here (setting $\mu=1$).}
\end{center}
\label{table:secondset}
\end{table}

%%%%%%%%%%%%%%%%%%%%%%%%TABLE%%%%%%%%%%%%%%%%%%%%%%%%%%%%%%%%%%%%%%%%

\end{document}